\documentclass[conference,a4paper]{IEEEtran}
\usepackage{xcolor}
\usepackage{balance}

\newcommand{\exportFigures}{false}
\usepackage{xcolor}

\usepackage{cite}
\usepackage{amsmath,amssymb,amsfonts,amsthm}
\usepackage{empheq}
\usepackage{algorithmic}

\usepackage{graphicx}

\usepackage{textcomp}

\usepackage{url}
\usepackage{bm}

\usepackage[caption=false,font=footnotesize]{subfig}

\usepackage[normalem]{ulem}

\usepackage{mathtools, cuted}

\usepackage{pdfpages} 

\usepackage{listings}

\usepackage{tikz}
\usetikzlibrary{angles}
\usetikzlibrary{arrows}
\usetikzlibrary{arrows.meta}
\usetikzlibrary{backgrounds}
\usetikzlibrary{calc}
\usetikzlibrary{decorations.pathmorphing}
\usetikzlibrary{fit}
\usetikzlibrary{patterns}
\usetikzlibrary{petri}
\usetikzlibrary{plotmarks}
\usetikzlibrary{positioning}
\usetikzlibrary{quotes}
\usetikzlibrary{shapes.misc}
\usetikzlibrary{spy}

\usepackage{pgfplots}
\pgfplotsset{compat=newest}
\usepgfplotslibrary{patchplots}
\usepgfplotslibrary{units}
\usepgfplotslibrary{polar}

\usepackage[most,theorems]{tcolorbox}
\tcbuselibrary{breakable}
\tcbset{every box/.style={enhanced,breakable}}

\usepackage[ruled,vlined]{algorithm2e}
\SetKwRepeat{Do}{do}{while}
\usepackage{setspace} 

\usepackage{cancel}

\usepackage{ifthen}

\newcommand{\tikzpng}[2]{
\ifthenelse{#1=1}
{\centering\input{#2.tex}}
{\centering\includegraphics[width=\figurewidth]{#2.png}}
}

\ifthenelse{\equal{\exportFigures}{true}}
{
  \usepgfplotslibrary{external}\tikzexternalize[prefix=tikz/]
}
{}

\newcommand{\ticked}{$\text{\rlap{$\checkmark$}}\square$}
\newcommand{\unticked}{{$\square$}}
\newcommand{\tick}[1]{\ifthenelse{#1=1}{\ticked}{\unticked}}

\newcounter{assump}

\definecolor{tw}{RGB}{51,183,150}

\newcommand{\vm}[1]{\ensuremath{\bm{#1}}} 

\DeclareMathOperator*{\argmax}{arg\,max}

\newlength{\figurewidth}
\newlength{\figureheight}

\ifCLASSOPTIONcompsoc
 \usepackage[caption=false,font=normalsize,labelfont=sf,textfont=sf]{subfig}
\else
 \usepackage[caption=false,font=footnotesize]{subfig}
\fi
\begin{document}
%
\title{Multipath-based Localization and Tracking considering Off-Body Channel Effects}

\author{\IEEEauthorblockN{
Thomas Wilding\IEEEauthorrefmark{4},   
Erik Leitinger\IEEEauthorrefmark{4},   
Klaus Witrisal\IEEEauthorrefmark{4}\IEEEauthorrefmark{1},    
}                                     
\thanks{This research was partly funded by the Austrian Research Promotion Agency (FFG) within the project UBSmart (project number: 859475). The financial support by the Christian Doppler Research Association, the Austrian Federal Ministry for Digital and Economic Affairs and the National Foundation for Research, Technology and Development is gratefully acknowledged.}
\IEEEauthorblockA{\IEEEauthorrefmark{4}
Graz University of Technology, Austria}
\IEEEauthorblockA{\IEEEauthorrefmark{1}
Christian Doppler Laboratory for Location-aware Electronic Systems}
\IEEEauthorblockA{ \emph{thomas.wilding@tugraz.at} }
}




\maketitle


\begin{abstract}
This paper deals with multipath-based positioning and tracking in off-body channels. 
An analysis of the effects introduced by the human body and the implications on positioning and tracking is presented based on channel measurements obtained in an indoor scenario. 
It shows the influence of the radio signal bandwidth on the human body induced field of view (FOV) and the number of multipath components (MPCs) detected and estimated by a deterministic maximum likelihood (ML) algorithm. 
A multipath-based positioning and tracking algorithm is proposed that associates these estimated MPC parameters with floor plan features and exploits a human body-dependent FOV function. The proposed algorithm is able to provide accurate position estimates even for an off-body radio channel in a multipath-prone environment with the signal bandwidth found to be a limiting factor.
\end{abstract}

\vskip0.5\baselineskip

%

%

\section{Introduction}\label{sec:introduction}
%

Robust radio-based localization has a varity of emerging applications. These applications include location-aware wireless communications services \cite{TarMupRauSloSveWym:SPM2014_LocationAware5G}, autonomous driving \cite{BressonTIV2017}, and the internet-of-things \cite{WinSPM2018}. 

Robust and accurate localization depends on many factors as for example on the hardware and the propagation environment. 
In fact, in multipath-prone environments it is especially challenging to achieve a predefined level of robustness. 
However, when designing robust localization algorithms the consideration of deteriorating effect on the radio signal caused by the human body itself is of great importance. 
The human body often leads to a complete blockage of line-of-sight paths to base stations and it induces dispersive effects on the radio signals.


\subsection{State-of-the-Art}

Multipath-based localization and tracking algorithms \cite{MeissnerWCL2014,LeitingerGNSS2016,WitrisalSPM2016, SchMarGenSanFie:AWPL2021_MultipathEnhancedLoc} and simultaneous localization and mapping (SLAM) algorithms \cite{LiJSTSP2013,GentnerTWC2016,LeitingerICC2017,LeitingerTWC2019, LeitingerICC2019,MendrzikJSTSP2019,LeitingerAsilomar2020_DataFusion} can leverage multiple propagation paths of radio signals for agent localization and can thus significantly improve localization accuracy and robustness. 
Usually, these algorithms model specular reflections of radio signals at flat
surfaces by virtual anchors (VAs) that are mirror images of the physical anchors (PAs) \cite{LeitingerJSAC2015,WitrisalWCL2016,hinteregger2016, MendrzikTWC2019}. 
Multipath-based SLAM algorithms detect and localize virtual anchors (VAs) and jointly estimate the time-varying position of mobile agents, whereas multipath-based localization algorithms assume prior knowledge of a floor plan to determine the locations of VAs. 

In this work, we assume the floor plan to be known and focus on the effect of the human body on multipath-based localization. 
Based on the off-body channel model introduced in \cite{WildingPIMRC2020,WildingEuCAP2021}, we extend the algorithm in \cite{MeissnerWCL2014} by considering the shadowing effect of the human body on specific multipath components (MPCs) and the LOS component. 
In particular, we introduce a human-body dependent field of view (FOV) function that directly considers the visibility of VAs in the data association (DA) between VAs and MPCs. 
With this, the number of possible VA-MPC association pairs is reduced, which significantly increases the robustness (and reduces the computational demand) of the optimal subpattern assignment (OSPA)-based DA method. 
Our main contributions can be summarized as follows:
\begin{itemize}
	\item We extend the algorithm in \cite{MeissnerWCL2014} by introducing a human body-dependent FOV function. 
	\item We show that the FOV function significantly increases the robustness of the OSPA-based DA.
	\item We demonstrate significant improvements in multipath-based localization and tracking performance using measured radio signals.
\end{itemize}


\section{System Model}\label{sec:system}


\begin{figure}[t]

\subfloat[floor plan]{\centering
\setlength{\figurewidth}{0.325\columnwidth}
\setlength{\figureheight}{0.325\columnwidth}
\def\datapath{./figures/m120_floorplan}
\input{./figures/m120_floorplan/m120_floorplan.tex}
\label{fig:scenario:floorplan}
}\hspace{6mm}
\subfloat[agent]{\centering
\includegraphics[width=0.35\columnwidth,trim=160 50 160 0,clip]{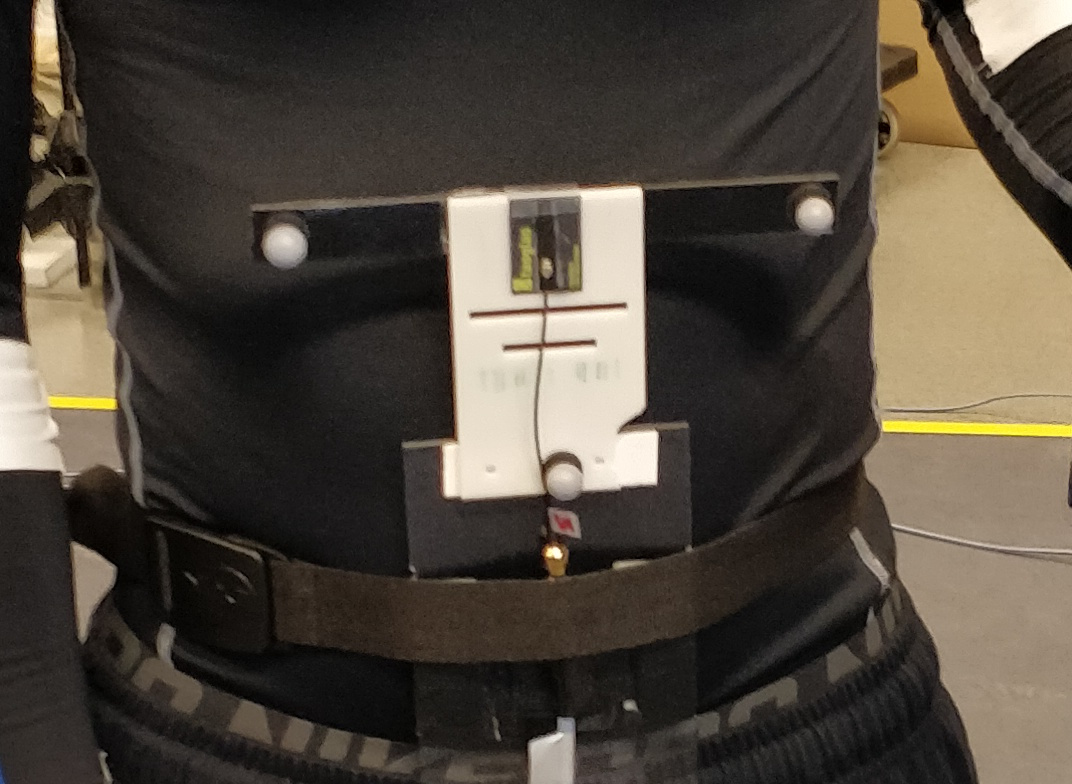}
\label{fig:scenario:agent}
}

\caption{Floor plan of the measurement environment showing the agent trajectory (gray) and the anchors $\vm{a}^{(j)}$ in {\protect\subref{fig:scenario:floorplan}} and the mobile device on the torso of a user together forming the agent in {\protect\subref{fig:scenario:agent}}.}\label{fig:scenario}
\end{figure}

The system under consideration contains a mobile agent and $J$ PAs at known fixed positions $\vm{a}^{(j)} = [x_\mathrm{a}^{(j)},y_\mathrm{a}^{(j)}]^\mathsf{T}\in\mathbb{R}^2$ with $j=1,\dots,J$.
At time step $n$, the agent is at position $\vm{p}_n = [x_n,y_n]^\mathsf{T}\in\mathbb{R}^2$ with the corresponding agent state vector $\vm{x}_n$ containing the agent position $\vm{p}_n$ and the agent velocity $\vm{v}_n$.

Considering off-body channels, the agent consists of two elements: a person that is moving through the environment, and a mobile device that is located at a fixed on-body position on that person (see Fig.~\ref{fig:scenario:agent}). 
With the mobile device being the position of interest, the agent position $\vm{p}_n$ is defined as the location of the mobile device.
Assuming an indoor environment, multipath propagation is modeled by an image source model with MPCs represented by VAs.
From available floorplan knowledge, the locations of the $K$ VAs are assumed known, with the position of the $k$th VA for the $j$th PA denoted as $\vm{a}_k^{(j)} = [x_{\mathrm{a},k}^{(j)},y_{\mathrm{a},k}^{(j)}]^\mathsf{T} \in\mathbb{R}^2$. 


At each time step $n$, $M_n^{(j)}$  distance measurements $d_{m,n}^{(j)}$ are extracted from signals exchanged between the $j$th anchor and the agent with $m=1,\dots,M_n^{(j)}$.
The model for the received signals is described in Section~\ref{sec:signal} alongside the estimation procedure to extract the distance estimates and further processing steps.
With each MPC associated with a corresponding VA, the distance measurement of the $k$th VA of the $j$th PA at time step $n$ is modeled as 
\begin{align}
z_{k,n}^{(j)} = d_{k,n}^{(j)} + e_k^{(j)} \label{eq:distance_measurement}
\end{align}
where $d_{k,n}^{(j)}=c\tau_{k,n}^{(j)} = \| \vm{p}_n - \vm{a}_k^{(j)} \|$ is the distance from agent to the VA and $e_{k,n}^{(j)}$ is zero mean Gaussian noise with standard deviation $\sigma_{k,n}^{(j)}$ and $\|\cdot\|$ is the vector norm.




\section{Signal Model and Processing}\label{sec:signal}

This section describes the received signal model for multipath channels in Sec.~\ref{sec:signal_model}, followed by the subsequent processing steps described applied to each signal snapshot in Sec.~\ref{sec:signal_model}.
For conciseness we omit the indices for time step $n$ and PA $j$ in the following sections, reintroducing them when necessary.

\subsection{Signal Model}\label{sec:signal_model}
Sampling the complex baseband signal $r(t)$ exchanged between the agent and a PA with sampling time $T_s$ yields the signal vector $\vm{r}=[r(0),r(T_s),\dots,r((N-1)T_s)]^\mathsf{T}\in\mathbb{C}^N$, described by a deterministic stochastic model as
\begin{align}
\vm{r} &= \sum_{k=1}^{K} \alpha_{k} \vm{s}(\tau_{k}) + \vm{r}_{\mathrm{s}} + \vm{w}\label{eq:signal}
\end{align}
where $\vm{s}(\tau) = [s(0-\tau),s(T_s-\tau),\dots,s((N-1)T_s-\tau)]^\mathsf{T}$ is the sampled $k$th MPC representing a delayed and scaled baseband pulse $s(t)$ and $\vm{w}$ is a vector containing AWGN samples with double sided power spectral density $N_0/2$ and variance $\sigma_{w}^2 = N_0/T_s$.
The deterministic part in \eqref{eq:signal} is the sum of $K$ MPCs carrying position information in form of the MPC delay $\tau_{k}$ representing the $k$th VA with complex valued MPC amplitude $\alpha_{k} \in \mathbb{C}$.

The stochastic part $\vm{r}_{\mathrm{s}}$ models stochastic interference via point source scattering \cite{SantosTWC2010a,SchubertTAP2013} which models scattering by a sum of stochastic MPCs as
\begin{align}
\vm{r}_{\mathrm{s}} = \sum_{\ell\in\mathcal{L}}\beta_{\ell} \vm{s}(\tau_{\ell})
\end{align}
where $\beta_{\ell}\in\mathbb{C}$ and $\tau_{\ell}$ are scattering coefficient and delay of the $\ell$th scattered component for $\mathcal{L}=1,\dots,{L}$ with random number of scattered components ${L}$. 
Scattering can be modeled in the spatial domain \cite{SchubertTAP2013,FlordelisTWC2020} or directly in the delay domain \cite{ErtelJSAC1999,JakobsenIZS2012}.
A scattering model was shown in \cite{WildingPIMRC2020} to be suitable to describe delay dispersive effects characteristic for off-body channels, with shadowing modeled in the MPC amplitudes and the distribution of the scattering coefficients.


\subsection{Processing Steps}\label{sec:processing}
At each time step $n$ a maximum likelihood (ML) based estimation procedure yields distance measurements used as input to a an OSPA-based DA algorithm which provides the necessary data for the tracking filter, described below.

\paragraph{Channel Estimation}
Delay measurements are obtained from the received signals $\vm{r}$ using a ML-based iterative estimator with a sparsity constraint on the MPC amplitudes. 
The estimator assumes an AWGN-only channel, with components attributable to the stochastic part in \eqref{eq:signal} treated as MPCs. 
The signal likelihood is denoted by 
\begin{align}
f(\vm{r};\vm{\tau},\vm{\alpha},\sigma_{w}^2) = (\pi\sigma_{w}^2)^{-N} \mathrm{exp}\big(\|\vm{r}-\vm{S}(\vm{\tau})\vm{\alpha}\|^2\sigma_{w}^{-2}\big)\label{eq:likelihood}
\end{align}
with $\vm{S} = [\vm{s}(\tau_1),\dots,\vm{s}(\tau_K)]$ and where the vectors $\vm{\tau}=[\tau_{1},\dots,\tau_{K}]^\mathsf{T}$ and $\vm{\alpha}=[\alpha_{1},\dots,\alpha_{K}]^\mathsf{T}$ contain the stacked delays and amplitudes of $K$ MPCs.
We use a deterministic ML estimator \cite{Ottersten1993} where the ML estimates for amplitudes $\hat{\vm{\alpha}}(\vm{\tau})$ and noise variance $\hat{\sigma}_w^2(\vm{\tau})$ obtained from \eqref{eq:likelihood} (see \cite{Ottersten1993}) are used to estimate the delays estimates $\hat{\vm{\tau}}$ by iteratively maximizing
\begin{align}
\hat{\vm{\tau}} = \argmax_{\vm{\tau}}~\mathrm{log}f(\vm{r};\vm{\tau},\hat{\vm{\alpha}}(\vm{\tau}),\hat{\sigma}_w^2(\vm{\tau})) + \lambda \|\hat{\vm{\alpha}}(\vm{\tau})\|_1
\end{align}
where $\|\cdot\|_1$ denotes the $\ell1$-norm, chosen as a heuristic penalty term to penalize closely spaced components.
The parameter $\lambda$ was chosen to fit the measurement data.
New components are added until the component signal-to-noise ratio (SNR) estimate falls below a threshold $\gamma$ \cite{NadlerTSP2011,LeitingerAsilomar2020}. The component SNR (neglecting overlap between MPCs) is defined as $\widehat{\mathrm{SNR}}_k = |\hat{\alpha}_k(\hat{\tau}_k)|^2\|\vm{s}(\hat{\tau}_k)\|/\hat{\sigma}_w^2(\hat{\vm{\tau}})$ where $\hat{\alpha}_k(\hat{\tau}_k)$ is the amplitude of the $k$th SMC from $\vm{\alpha}(\hat{\vm{\tau}})$.

Reintroducing the indices for time steps and PAs $M_n^{(j)} = \hat{K}_n^{(j)}$ delay measurements $\vm{z}_n^{(j)} = c\hat{\vm{\tau}}_n^{(j)}$ are obtained at each time step, containing clutter measurements and measurements attributable to VAs by \eqref{eq:distance_measurement}.

\paragraph{Data Association (DA)}
Distance measurements $\vm{z}_n^{(j)}$ are associated to candidate delays of known VAs by snapshot level DA based on the OSPA approach \cite{SchuhmacherTSP2008,MeissnerWCL2014}.
The candidate delays are calculated from the set of VAs that are visible at an agent position hypothesis $\hat{\vm{p}}_n$ at time step $n$, with the FOV of the agent included in the DA by only allowing PAs and VAs that are within the FOV to be used as DA candidates.
The evaluate the visibility of VAs or PAs, the orientation of the agent is assumed known and to allow investigating the improvements attributable to multipath assistance. 
Note that in a realistic setting the orientation needs to be estimated or obtained via additional sensors. 
The FOV is further discussed in Sec.~\ref{sec:results:tracking}.

The DA step yields a subset $\tilde{\vm{z}}_n^{(j)}$ of measurements from $\vm{z}_n^{(j)}$ that are associated to VAs or PAs and a DA vector to assigning a unique VA or PA to each element in $\tilde{\vm{z}}_n^{(j)}$.
A cut-off distance $d_c$ ensures that only associations within a chosen deviation between estimated delays and predicted delays are possible.


\paragraph{EKF-based Tracking Filter}
The associated measurements $\tilde{\vm{z}}_n^{(j)}$ from the DA step are used to track the agent position by an extended Kalman filter (EKF) similar to the one described in \cite{MeissnerWCL2014}.
For each measurement in $\tilde{\vm{z}}_n^{(j)}$ the measurement uncertainty for the EKF is computed from the AWGN CRLB as
\begin{align}
\sigma_k^{(j)} \triangleq \sqrt{8\pi^2{{\beta}}^2 \widehat{\mathrm{SNR}}_k^{(j)}} \label{eq:uncertainty}
\end{align}
where $\beta^2$ is the mean squared signal bandwidth \cite{hinteregger2016}.
To improve the performance, $\beta$ is reduced by a factor $\rho$ when computing \eqref{eq:uncertainty}, i.e., inserting $\beta^2/\rho^2$ for $\beta^2$.

\section{Results}\label{sec:results}

\subsection{Measurement Description}\label{sec:results:measurements}

\begin{figure}[t!]
	\centering
	
	\subfloat[anchor $\vm{a}^{(1)}$]{\hspace{-5mm}
		\setlength{\figurewidth}{0.37\columnwidth}
		\setlength{\figureheight}{0.25\columnwidth}
		\def\datapath{./figures/m120_K_CDF}
%
%
 
\pgfplotsset{every axis/.append style={
  label style={font=\footnotesize},
  legend style={font=\footnotesize},
  tick label style={font=\footnotesize},
  xticklabel={
    \ifdim \tick pt < 0pt
      \pgfmathparse{abs(\tick)}%
      \llap{$-{}$}\pgfmathprintnumber{\pgfmathresult}
   \else
      \pgfmathprintnumber{\tick}
   \fi
}}}
 
\begin{tikzpicture}

\begin{axis}[%
width=0.9\figurewidth,
height=\figureheight,
at={(0\figurewidth,0\figureheight)},
scale only axis,
xmin=0,
xmax=25,
xlabel={number of SMCs},
ymin=0,
ymax=1,
ylabel={CDF},
xmajorgrids,
ymajorgrids,
legend style={at={(1,1.01)}, legend columns=1, anchor=north west, legend cell align=left, align=left, draw=none,fill=none},
grid style=dotted,
major tick length=1mm,
clip mode=individual,
xlabel style = {yshift=1.5mm},
ylabel style = {yshift=-1mm},
xtick={0,5,...,30},
]

\addplot [color=red, densely dotted, line width=1pt]
  table[]{\datapath/m120_fc6950MHz_bw0500MHz_K_CDF-5.tsv};
\addlegendentry{geom.}

\addplot [color=black,mark=triangle,mark options={solid,black},mark repeat=5,mark phase=5, densely dashed, line width=0.5pt,forget plot]
  table[]{\datapath/m120_fc6950MHz_bw3939MHz_K_CDF-1.tsv};

\addplot [color=black,mark=triangle,mark options={solid,black},mark repeat=3,mark phase=5, line width=0.5pt]
  table[]{\datapath/m120_fc6950MHz_bw3939MHz_K_CDF-2.tsv};
\addlegendentry{$4~\mathrm{GHz}$}

%

\addplot [color=green!50!black,mark=o, mark options={solid,green!50!black},mark repeat=5,mark phase=7,dashed, line width=0.5pt,forget plot]
  table[]{\datapath/m120_fc6950MHz_bw2000MHz_K_CDF-1.tsv};

\addplot [color=green!50!black, mark=o, mark options={solid,green!50!black},mark repeat=3,mark phase=7,line width=0.5pt]
  table[]{\datapath/m120_fc6950MHz_bw2000MHz_K_CDF-2.tsv};
\addlegendentry{$2~\mathrm{GHz}$}

%

%
%
%

\addplot [color=violet,densely dashed, line width=0.5pt,forget plot,mark=none, mark options={solid,violet},mark repeat=4,mark phase=3]
  table[]{\datapath/m120_fc6950MHz_bw0500MHz_K_CDF-1.tsv};

\addplot [color=violet, line width=0.5pt,mark=none, mark options={solid,violet},mark repeat=4,mark phase=3]
  table[]{\datapath/m120_fc6950MHz_bw0500MHz_K_CDF-2.tsv};
\addlegendentry{$500~\mathrm{MHz}$}

%

\legend{}

\end{axis}
\end{tikzpicture}%
		\label{fig:K_CDF:A1}
	}
	\subfloat[anchor $\vm{a}^{(2)}$]{\hspace{-4mm}
		\setlength{\figurewidth}{0.37\columnwidth}
		\setlength{\figureheight}{0.25\columnwidth}
		\def\datapath{./figures/m120_K_CDF}
%
%
 
\pgfplotsset{every axis/.append style={
  label style={font=\footnotesize},
  legend style={font=\footnotesize},
  tick label style={font=\footnotesize},
  xticklabel={
    \ifdim \tick pt < 0pt
      \pgfmathparse{abs(\tick)}%
      \llap{$-{}$}\pgfmathprintnumber{\pgfmathresult}
   \else
      \pgfmathprintnumber{\tick}
   \fi
}}}
 
\begin{tikzpicture}

\begin{axis}[%
width=0.9\figurewidth,
height=\figureheight,
at={(0\figurewidth,0\figureheight)},
scale only axis,
xmin=0,
xmax=25,
xlabel={number of SMCs},
ymin=0,
ymax=1,
ylabel={CDF},
xmajorgrids,
ymajorgrids,
legend style={at={(1,1.01)}, legend columns=4, anchor=north west, legend cell align=left, align=left, draw=none,fill=none},
legend to name = cdfs,
grid style=dotted,
major tick length=1mm,
clip mode=individual,
xlabel style = {yshift=1.5mm},
ylabel style = {yshift=-1mm},
xtick={0,5,...,30},
]

\addplot [color=red, densely dotted, line width=1pt]
  table[]{\datapath/m120_fc6950MHz_bw0500MHz_K_CDF-6.tsv};
\addlegendentry{geom.}

%

\addplot [color=black,mark=triangle,mark options={solid,black},mark repeat=5,mark phase=4, densely dashed, line width=0.5pt,forget plot]
  table[]{\datapath/m120_fc6950MHz_bw3939MHz_K_CDF-3.tsv};

\addplot [color=black,mark=triangle,mark options={solid,black},mark repeat=4,mark phase=5, line width=0.5pt]
  table[]{\datapath/m120_fc6950MHz_bw3939MHz_K_CDF-4.tsv};
\addlegendentry{$4~\mathrm{GHz}$}

%

\addplot [color=green!50!black,mark=o, mark options={solid,green!50!black},mark repeat=5,mark phase=6, densely dashed, line width=0.5pt,forget plot]
  table[]{\datapath/m120_fc6950MHz_bw2000MHz_K_CDF-3.tsv};

\addplot [color=green!50!black,mark=o, mark options={solid,green!50!black},mark repeat=5,mark phase=3, line width=0.5pt]
  table[]{\datapath/m120_fc6950MHz_bw2000MHz_K_CDF-4.tsv};
\addlegendentry{$2~\mathrm{GHz}$}

%
%
%

%

\addplot [color=violet, densely dashed, line width=0.5pt,forget plot,mark=none, mark options={solid,violet},mark repeat=4,mark phase=3]
  table[]{\datapath/m120_fc6950MHz_bw0500MHz_K_CDF-3.tsv};

\addplot [color=violet, line width=0.5pt,mark=none, mark options={solid,violet},mark repeat=4,mark phase=3]
  table[]{\datapath/m120_fc6950MHz_bw0500MHz_K_CDF-4.tsv};
\addlegendentry{$500~\mathrm{MHz}$}

\end{axis}
\end{tikzpicture}%
		\label{fig:K_CDF:A2}
	}
	
	\ref{cdfs}
	
	\caption{CDF for number of estimated (dashed) and associated (solid) MPCs for each anchor $\vm{a}^{(j)}$ for bandwidth of $BW = \{500~\mathrm{MHz},2~\mathrm{GHz},4~\mathrm{GHz}\}$.}\label{fig:K_CDF}
\end{figure}
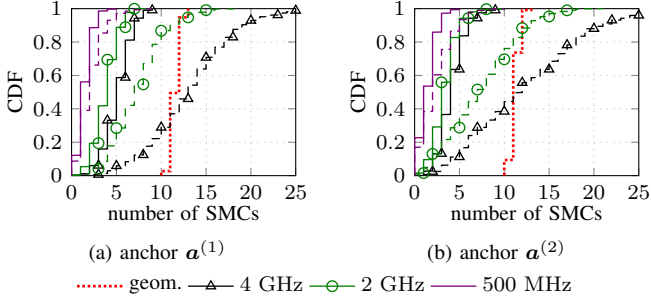

Channel measurements were performed with a correlative channel sounder covering a frequency range of $f = 3.1-10.6~\mathrm{GHz}$.
Along a trajectory in a laboratory environment consisting of $P=900$ positions, channel responses to $J=2$ PAs were recorded at each trajectory position, i.e., for time steps $n=1,\dots,P$. 
The ground truth position $\vm{p}_{\mathrm{gt},n}$ and orientation $\psi_{\mathrm{gt},n}$ of the agent were obtained form an optical tracking system (OTS), alongside the positions of the PAs.
The channel sounder was calibrated up to but not including the antennas, the selection of the measurement bandwidth was performed in post-processing. 
All measurements were fully synchronized.
The necessary VAs for both PAs were computed geometrically up to second order from an available floor plan shown in Fig.~\ref{fig:scenario:floorplan}.
To improve the accuracy of the floor plan the geometric VAs were refined by grid based ML estimates of each VA position within a $1\times 1~\mathrm{m}$ region centered at the initial VA location \cite{MeissnerWCL2014}, using the signals along the full trajectory and the maximum bandwidth of the channel sounder. 
The agent consisted of a mobile device equipped with a single antenna directly positioned at the center torso of a person as shown in Fig.~\ref{fig:scenario:agent}.

%
%


\subsection{Measurement Analysis}\label{sec:results:analysis}



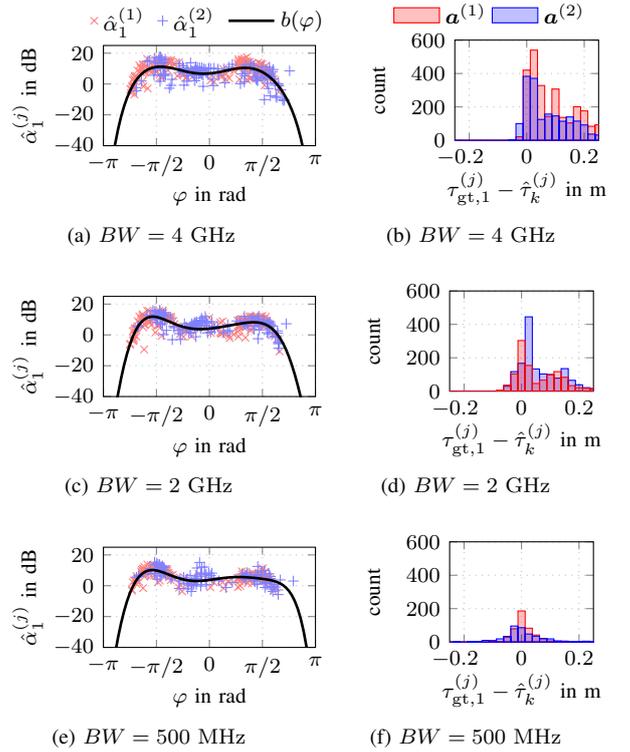
\begin{figure}[t!]
	
	\subfloat[$BW=4~\mathrm{GHz}$]{
		\setlength{\figurewidth}{0.35\columnwidth}
		\setlength{\figureheight}{0.15\columnwidth}
		\def\datapath{./figures/m120_fc6950MHz_bw3939MHz_ampPL}
%
%
 
\pgfplotsset{every axis/.append style={
  label style={font=\footnotesize},
  legend style={font=\footnotesize},
  tick label style={font=\footnotesize},
  xticklabel={
    \ifdim \tick pt < 0pt
      \pgfmathparse{abs(\tick)}%
      \llap{$-{}$}\pgfmathprintnumber{\pgfmathresult}
   \else
      \pgfmathprintnumber{\tick}
   \fi
}}}
 
\begin{tikzpicture}

\begin{axis}[%
width=0.9\figurewidth,
height=\figureheight,
at={(0\figurewidth,0\figureheight)},
scale only axis,
xmin=-pi,
xmax=pi,
xlabel={$\varphi$ in rad},
xtick={-pi,-pi/2,0,pi/2,pi},
xticklabels={$-\pi$,$-\pi/2$,$0$,$\pi/2$,$\pi$},
ymin=-40,
ymax=25,
ylabel={$\hat{\alpha}_1^{(j)}$ in dB},
axis background/.style={fill=white},
xmajorgrids,
ymajorgrids,
legend style={at={(0.5,1)},anchor=south,legend cell align=left, legend columns=3, align=left, draw=none,fill=none},
grid style=dotted,
major tick length=1mm,
clip mode=individual,
xlabel style = {yshift=1mm},
ylabel style = {yshift=-1mm},
]

\addplot [color=red, draw=none, mark=x, mark options={solid, red!50!white},only marks,mark size=2pt,each nth point = {5}]
  table[]{\datapath/m120_fc6950MHz_bw3939MHz_ampPL-1.tsv};
\addlegendentry{$\hat{\alpha}_1^{(1)}$  }

\addplot [color=blue, draw=none, mark=+, mark options={solid, blue!50!white},only marks,mark size=2pt,each nth point = {5}]
  table[]{\datapath/m120_fc6950MHz_bw3939MHz_ampPL-2.tsv};
\addlegendentry{$\hat{\alpha}_1^{(2)}$  }

\addplot [color=black, line width=1pt]
  table[]{\datapath/m120_fc6950MHz_bw3939MHz_ampPL-3.tsv};
\addlegendentry{$b(\varphi)$}


\end{axis}
\end{tikzpicture}%
		\label{fig:LOS:amp:4GHz}
	}\hspace{-5mm}
	\subfloat[$BW=4~\mathrm{GHz}$]{
		\setlength{\figurewidth}{0.22\columnwidth}
		\setlength{\figureheight}{0.15\columnwidth}
		\def\datapath{./figures/m120_fc6950MHz_bw3939MHz_LOS_error_hist}

%
%
 
\pgfplotsset{every axis/.append style={
  label style={font=\footnotesize},
  legend style={font=\footnotesize},
  tick label style={font=\footnotesize},
  xticklabel={
    \ifdim \tick pt < 0pt
      \pgfmathparse{abs(\tick)}%
      \llap{$-{}$}\pgfmathprintnumber{\pgfmathresult}
   \else
      \pgfmathprintnumber{\tick}
   \fi
}}}
 
\begin{tikzpicture}

\begin{axis}[%
width=0.971\figurewidth,
height=\figureheight,
at={(0\figurewidth,0\figureheight)},
scale only axis,
xmin=-0.25,
xmax=0.25,
xlabel={$\tau_{\mathrm{gt},1}^{(j)}-\hat{\tau}_k^{(j)}$ in m},
ymin=0,
ymax=600,
ylabel={count},
axis background/.style={fill=white},
xmajorgrids,
ymajorgrids,
legend style={at={(1,1.02)},legend columns=2, align=left, align=left, draw=none,fill=none,anchor=south east},
grid style=dotted,
major tick length=1mm,
clip mode=individual,
xlabel style = {yshift=1.5mm},
ylabel style = {yshift=1mm},
]
\addplot[ybar interval, fill=red, fill opacity=0.25, draw=red, area legend] table[] {\datapath/m120_fc6950MHz_bw3939MHz_LOS_error_hist-1.tsv};
\addlegendentry{$\vm{a}^{(1)}$}

\addplot[ybar interval, fill=blue, fill opacity=0.25, draw=blue, area legend] table[] {\datapath/m120_fc6950MHz_bw3939MHz_LOS_error_hist-2.tsv};
\addlegendentry{$\vm{a}^{(2)}$}


\end{axis}
\end{tikzpicture}%
		\label{fig:LOS:error:4GHz}
	}

	\subfloat[$BW=2~\mathrm{GHz}$]{
		\setlength{\figurewidth}{0.35\columnwidth}
		\setlength{\figureheight}{0.15\columnwidth}
		\def\datapath{./figures/m120_fc6950MHz_bw2000MHz_ampPL}
%
%
 
\pgfplotsset{every axis/.append style={
  label style={font=\footnotesize},
  legend style={font=\footnotesize},
  tick label style={font=\footnotesize},
  xticklabel={
    \ifdim \tick pt < 0pt
      \pgfmathparse{abs(\tick)}%
      \llap{$-{}$}\pgfmathprintnumber{\pgfmathresult}
   \else
      \pgfmathprintnumber{\tick}
   \fi
}}}
 
\begin{tikzpicture}

\begin{axis}[%
width=0.9\figurewidth,
height=\figureheight,
at={(0\figurewidth,0\figureheight)},
scale only axis,
xmin=-pi,
xmax=pi,
xlabel={$\varphi$ in rad},
xtick={-pi,-pi/2,0,pi/2,pi},
xticklabels={$-\pi$,$-\pi/2$,$0$,$\pi/2$,$\pi$},
ymin=-40,
ymax=25,
ylabel={$\hat{\alpha}_1^{(j)}$ in dB},
axis background/.style={fill=white},
xmajorgrids,
ymajorgrids,
legend style={at={(0.5,1)},anchor=south,legend cell align=left, legend columns=3, align=left, draw=none,fill=none},
grid style=dotted,
major tick length=1mm,
clip mode=individual,
xlabel style = {yshift=1mm},
ylabel style = {yshift=-1mm},
]

\addplot [color=red, draw=none, mark=x, mark options={solid, red!50!white},only marks,mark size=2pt,each nth point = {5}]
  table[]{\datapath/m120_fc6950MHz_bw2000MHz_ampPL-1.tsv};
\addlegendentry{$\hat{\alpha}_1^{(1)}$  }

\addplot [color=blue, draw=none, mark=+, mark options={solid, blue!50!white},only marks,mark size=2pt,each nth point = {5}]
  table[]{\datapath/m120_fc6950MHz_bw2000MHz_ampPL-2.tsv};
\addlegendentry{$\hat{\alpha}_1^{(2)}$  }

\addplot [color=black, line width=1pt]
  table[]{\datapath/m120_fc6950MHz_bw2000MHz_ampPL-3.tsv};
\addlegendentry{$b(\varphi)$}

%
%

\legend{}

\end{axis}
\end{tikzpicture}%
		\label{fig:LOS:amp:2GHz}
	}\hspace{-5mm}
	\subfloat[$BW=2~\mathrm{GHz}$]{
		\setlength{\figurewidth}{0.22\columnwidth}
		\setlength{\figureheight}{0.15\columnwidth}
		\def\datapath{./figures/m120_fc6950MHz_bw2000MHz_LOS_error_hist}
%
%
 
\pgfplotsset{every axis/.append style={
  label style={font=\footnotesize},
  legend style={font=\footnotesize},
  tick label style={font=\footnotesize},
  xticklabel={
    \ifdim \tick pt < 0pt
      \pgfmathparse{abs(\tick)}%
      \llap{$-{}$}\pgfmathprintnumber{\pgfmathresult}
   \else
      \pgfmathprintnumber{\tick}
   \fi
}}}
 
\begin{tikzpicture}

\begin{axis}[%
width=0.971\figurewidth,
height=\figureheight,
at={(0\figurewidth,0\figureheight)},
scale only axis,
xmin=-0.25,
xmax=0.25,
xlabel={$\tau_{\mathrm{gt},1}^{(j)}-\hat{\tau}_k^{(j)}$ in m},
ymin=0,
ymax=600,
ylabel={count},
axis background/.style={fill=white},
xmajorgrids,
ymajorgrids,
legend style={at={(0,1.02)},legend cell,legend columns=2, align=left, align=left, draw=none,fill=none,anchor=south west},
grid style=dotted,
major tick length=1mm,
clip mode=individual,
xlabel style = {yshift=1.5mm},
ylabel style = {yshift=1mm},
]
\addplot[ybar interval, fill=blue, fill opacity=0.25, draw=blue, area legend] table[] {\datapath/m120_fc6950MHz_bw2000MHz_LOS_error_hist-1.tsv};
\addlegendentry{$\vm{a}^{(1)}$}

\addplot[ybar interval, fill=red, fill opacity=0.25, draw=red, area legend] table[] {\datapath/m120_fc6950MHz_bw2000MHz_LOS_error_hist-2.tsv};
\addlegendentry{$\vm{a}^{(2)}$}

\legend{}

\end{axis}
\end{tikzpicture}%
		\label{fig:LOS:error:2GHz}
	}

	\subfloat[$BW=500~\mathrm{MHz}$]{
		\setlength{\figurewidth}{0.35\columnwidth}
		\setlength{\figureheight}{0.15\columnwidth}
		\def\datapath{./figures/m120_fc6950MHz_bw0500MHz_ampPL}
%
%
 
\pgfplotsset{every axis/.append style={
  label style={font=\footnotesize},
  legend style={font=\footnotesize},
  tick label style={font=\footnotesize},
  xticklabel={
    \ifdim \tick pt < 0pt
      \pgfmathparse{abs(\tick)}%
      \llap{$-{}$}\pgfmathprintnumber{\pgfmathresult}
   \else
      \pgfmathprintnumber{\tick}
   \fi
}}}
 
\begin{tikzpicture}

\begin{axis}[%
width=0.9\figurewidth,
height=\figureheight,
at={(0\figurewidth,0\figureheight)},
scale only axis,
xmin=-pi,
xmax=pi,
xlabel={$\varphi$ in rad},
xtick={-pi,-pi/2,0,pi/2,pi},
xticklabels={$-\pi$,$-\pi/2$,$0$,$\pi/2$,$\pi$},
ymin=-40,
ymax=25,
ylabel={$\hat{\alpha}_1^{(j)}$ in dB},
axis background/.style={fill=white},
xmajorgrids,
ymajorgrids,
legend style={at={(0.5,1)},anchor=south,legend cell align=left, legend columns=3, align=left, draw=none,fill=none},
grid style=dotted,
major tick length=1mm,
clip mode=individual,
xlabel style = {yshift=1mm},
ylabel style = {yshift=-1mm},
]

\addplot [color=red, draw=none, mark=x, mark options={solid, red!50!white},only marks,mark size=2pt,each nth point = {5}]
  table[]{\datapath/m120_fc6950MHz_bw0500MHz_ampPL-1.tsv};
\addlegendentry{$\hat{\alpha}_1^{(1)}$  }

\addplot [color=blue, draw=none, mark=+, mark options={solid, blue!50!white},only marks,mark size=2pt,each nth point = {5}]
  table[]{\datapath/m120_fc6950MHz_bw0500MHz_ampPL-2.tsv};
\addlegendentry{$\hat{\alpha}_1^{(2)}$  }

\addplot [color=black, line width=1pt]
  table[]{\datapath/m120_fc6950MHz_bw0500MHz_ampPL-3.tsv};
\addlegendentry{$b(\varphi)$}

%
%

\legend{}

\end{axis}
\end{tikzpicture}%
		\label{fig:LOS:amp:500MHz}
	}\hspace{-5mm}
	\subfloat[$BW=500~\mathrm{MHz}$]{
		\setlength{\figurewidth}{0.22\columnwidth}
		\setlength{\figureheight}{0.15\columnwidth}
		\def\datapath{./figures/m120_fc6950MHz_bw0500MHz_LOS_error_hist}
%
%
 
\pgfplotsset{every axis/.append style={
  label style={font=\footnotesize},
  legend style={font=\footnotesize},
  tick label style={font=\footnotesize},
  xticklabel={
    \ifdim \tick pt < 0pt
      \pgfmathparse{abs(\tick)}%
      \llap{$-{}$}\pgfmathprintnumber{\pgfmathresult}
   \else
      \pgfmathprintnumber{\tick}
   \fi
}}}
 
\begin{tikzpicture}

\begin{axis}[%
width=0.971\figurewidth,
height=\figureheight,
at={(0\figurewidth,0\figureheight)},
scale only axis,
xmin=-0.25,
xmax=0.25,
xlabel={$\tau_{\mathrm{gt},1}^{(j)}-\hat{\tau}_k^{(j)}$ in m},
ymin=0,
ymax=600,
ylabel={count},
axis background/.style={fill=white},
xmajorgrids,
ymajorgrids,
legend style={at={(0,1.02)},legend cell,legend columns=2, align=left, align=left, draw=none,fill=none,anchor=south west},
grid style=dotted,
major tick length=1mm,
clip mode=individual,
xlabel style = {yshift=1.5mm},
ylabel style = {yshift=1mm},
]
\addplot[ybar interval, fill=red, fill opacity=0.25, draw=red, area legend] table[] {\datapath/m120_fc6950MHz_bw0500MHz_LOS_error_hist-1.tsv};
\addlegendentry{$\vm{a}^{(1)}$}

\addplot[ybar interval, fill=blue, fill opacity=0.25, draw=blue, area legend] table[] {\datapath/m120_fc6950MHz_bw0500MHz_LOS_error_hist-2.tsv};
\addlegendentry{$\vm{a}^{(2)}$}

\legend{}

\end{axis}
\end{tikzpicture}%
		\label{fig:LOS:error:500MHz}
	}

	\caption{Associated LOS amplitudes over AOA $\varphi$ at the agent for both anchors compensated for free space path loss in {\protect\subref{fig:LOS:amp:4GHz}, \protect\subref{fig:LOS:amp:2GHz} and \protect\subref{fig:LOS:amp:500MHz}} and delay error histogram for all SMC estimates $\hat{\tau}_k^{(j)}$ relative to the ground truth LOS delay $\tau_{\mathrm{gt},1}^{(j)}$ in {\protect\subref{fig:LOS:error:4GHz}, \protect\subref{fig:LOS:error:2GHz} and \protect\subref{fig:LOS:error:500MHz}. Results are shown for a bandwidth of $BW = \{500~\mathrm{MHz},2~\mathrm{GHz},4~\mathrm{GHz}\}$.}
	}\label{fig:LOS}
\end{figure}

\begin{figure*}[t!]
	\centering
	
	\subfloat[CDF]{
		\setlength{\figurewidth}{0.4\columnwidth}
		\setlength{\figureheight}{0.25\columnwidth}
		\def\datapath{./figures/m120_cdf}
%
%
 
\pgfplotsset{every axis/.append style={
  label style={font=\footnotesize},
  legend style={font=\footnotesize},
  tick label style={font=\footnotesize},
  xticklabel={
    \ifdim \tick pt < 0pt
      \pgfmathparse{abs(\tick)}%
      \llap{$-{}$}\pgfmathprintnumber{\pgfmathresult}
   \else
      \pgfmathprintnumber{\tick}
   \fi
}}}
 
\begin{tikzpicture}

\begin{axis}[%
width=0.951\figurewidth,
height=\figureheight,
at={(0\figurewidth,0\figureheight)},
scale only axis,
xmin=0,
xmax=1,
xlabel={position error in m},
ymin=0,
ymax=1,
ylabel={CDF},
axis background/.style={fill=white},
xmajorgrids,
ymajorgrids,
legend style={legend cell align=left, align=left, draw=none, fill=none},
grid style=dotted,
major tick length=1mm,
clip mode=individual,
xlabel style = {yshift=1mm},
ylabel style = {yshift=-1mm},
]

\addplot [color=green!50!black,densely dashed,line width=0.5pt,mark=o,mark options={solid,green!50!black},mark repeat=100,mark phase=40]
  table[]{\datapath/m120_cdf-1.tsv};
\addlegendentry{$BW=2~\mathrm{GHz}$ (III)}

\addplot [color=green!50!black,solid,line width=0.5pt,mark=o,mark options={solid,green!50!black},mark repeat=100,mark phase=40]
  table[]{\datapath/m120_cdf-2.tsv};
\addlegendentry{$BW=2~\mathrm{GHz}$ (II)}

\addplot [color=green!50!black,dashdotted,line width=0.5pt,mark=o,mark options={solid,green!50!black},mark repeat=100,mark phase=40]
  table[]{\datapath/m120_cdf-3.tsv};
\addlegendentry{$BW=2~\mathrm{GHz}$ (I)}

\addplot [color=black, densely dashed, line width=0.5pt,mark=triangle,mark options={solid,black},mark repeat=50,mark phase=60]
  table[]{\datapath/m120_cdf-4.tsv};
\addlegendentry{$BW=4~\mathrm{GHz}$ (III)}

\addplot [color=black, solid, line width=0.5pt,mark=triangle,mark options={solid,black},mark repeat=50,mark phase=60]
  table[]{\datapath/m120_cdf-5.tsv};
\addlegendentry{$BW=4~\mathrm{GHz}$ (II)}

\addplot [color=black, dashdotted, line width=0.5pt,mark=triangle,mark options={solid,black},mark repeat=50,mark phase=60]
  table[]{\datapath/m120_cdf-6.tsv};
\addlegendentry{$BW=4~\mathrm{GHz}$ (I) }

\addplot [color=violet,densely dashed, line width=0.5pt]
  table[]{\datapath/m120_cdf-7.tsv};
\addlegendentry{$BW=500~\mathrm{MHz}$ (III)}

\addplot [color=violet,solid, line width=0.5pt]
  table[]{\datapath/m120_cdf-8.tsv};
\addlegendentry{$BW=500~\mathrm{MHz}$ (II)}

\addplot [color=violet,dashdotted, line width=0.5pt]
  table[]{\datapath/m120_cdf-9.tsv};
\addlegendentry{$BW=500~\mathrm{MHz}$ (I)}

%
%
%
%
%
%
%
%
%
%
%
%

\legend{}

\end{axis}
\end{tikzpicture}%
		\label{fig:EKF:CDF}
	}
	\subfloat[position error]{
		\setlength{\figurewidth}{1.2\columnwidth}
		\setlength{\figureheight}{0.25\columnwidth}
		\def\datapath{./figures/m120_error}
%
%
 
\pgfplotsset{every axis/.append style={
  label style={font=\footnotesize},
  legend style={font=\footnotesize},
  tick label style={font=\footnotesize},
  xticklabel={
    \ifdim \tick pt < 0pt
      \pgfmathparse{abs(\tick)}%
      \llap{$-{}$}\pgfmathprintnumber{\pgfmathresult}
   \else
      \pgfmathprintnumber{\tick}
   \fi
}}}
 
\begin{tikzpicture}

\begin{axis}[%
width=0.951\figurewidth,
height=\figureheight,
at={(0\figurewidth,0\figureheight)},
scale only axis,
unbounded coords=jump,
xmin=0,
xmax=900,
xlabel={time step $n$},
ymin=0,
ymax=1,
ylabel={position error in m},
axis background/.style={fill=white},
xmajorgrids,
ymajorgrids,
legend style={legend cell align=left, align=left, draw=none, fill=none,legend columns=5},
legend to name = tracks,
grid style=dotted,
major tick length=1mm,
clip mode=individual,
xlabel style = {yshift=1mm},
ylabel style = {yshift=-1mm},
]

\addplot [color=green!50!black,densely dashed,line width=0.5pt,mark=o,mark options={solid,green!50!black},mark repeat=100,mark phase=40]
  table[]{\datapath/m120_error-1.tsv};
\addlegendentry{$BW=2~\mathrm{GHz}$ (III)}

\addplot [color=green!50!black,solid,line width=0.5pt,mark=o,mark options={solid,green!50!black},mark repeat=100,mark phase=40]
  table[]{\datapath/m120_error-2.tsv};
\addlegendentry{$BW=2~\mathrm{GHz}$ (II)}

\addplot [color=green!50!black,dashdotted,line width=0.5pt,mark=o,mark options={solid,green!50!black},mark repeat=100,mark phase=40]
  table[]{\datapath/m120_error-3.tsv};
\addlegendentry{$BW=2~\mathrm{GHz}$ (I)}

\addplot [color=black, densely dashed, line width=0.5pt,mark=triangle,mark options={solid,black},mark repeat=50,mark phase=60]
  table[]{\datapath/m120_error-4.tsv};
\addlegendentry{$BW=4~\mathrm{GHz}$ (III)}

\addplot [color=black, solid, line width=0.5pt,mark=triangle,mark options={solid,black},mark repeat=50,mark phase=60]
  table[]{\datapath/m120_error-5.tsv};
\addlegendentry{$BW=4~\mathrm{GHz}$ (II)}

\addplot [color=black, dashdotted, line width=0.5pt,mark=triangle,mark options={solid,black},mark repeat=50,mark phase=60]
  table[]{\datapath/m120_error-6.tsv};
\addlegendentry{$BW=4~\mathrm{GHz}$ (I) }

\addplot [color=violet,densely dashed, line width=0.5pt]
  table[]{\datapath/m120_error-7.tsv};
\addlegendentry{$BW=500~\mathrm{MHz}$ (III)}

\addplot [color=violet,solid, line width=0.5pt]
  table[]{\datapath/m120_error-8.tsv};
\addlegendentry{$BW=500~\mathrm{MHz}$ (II)}

\addplot [color=violet,dashdotted, line width=0.5pt]
  table[]{\datapath/m120_error-9.tsv};
\addlegendentry{$BW=500~\mathrm{MHz}$ (I)}

\end{axis}
\end{tikzpicture}%
		\label{fig:EKF:error}
	}
	
	\ref{tracks}
	\caption{CDF for position error using ground truth orientation for multipath-based tracking (solid) and LOS-only tracking (dashed) for varying bandwidth of $BW = \{500~\mathrm{MHz},2~\mathrm{GHz},4~\mathrm{GHz}\}$ shown in {\protect\subref{fig:EKF:CDF}}, the corresponding position error along the full trajectory containing $P=900$ points is shown in {\protect\subref{fig:EKF:error}}. The three configuration are indicated as (I) LOS-only with FOV, (II) multipath-based with FOV and (III) multipath-based without FOV.}
	\label{fig:EKF}
\end{figure*}
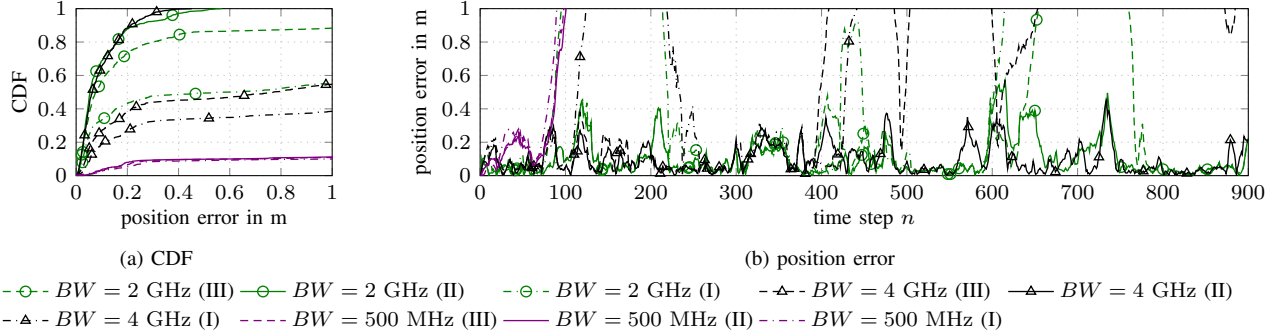

Using distance measurements obtained by the channel estimator described in Sec.~\ref{sec:processing}, the number of visible VAs and the LOS error along the trajectory are investigated by performing the DA for each time step with ground truth agent position $\vm{p}_{\mathrm{gt},n}$. 
This analysis was performed for varying bandwidth values $BW=\{4\,\mathrm{GHz},2\,\mathrm{GHz},500\,\mathrm{MHz}\}$ with the SNR threshold set to $\gamma=8~\mathrm{dB}$.

Figure~\ref{fig:K_CDF} shows CDFs of the number of measurements (dashed) and measurements associated to a VA or PA (solid) for both PAs for the full trajectory.
The figures include the CDFs for the number of components that would be visible geometrically (red dotted). 
While the number of distance measurements can generally be expected to be larger than the number of associated measurements, both the number of measurements and associated measurements reduces with bandwidth for the used estimator.
Note that for $BW=\{4\,\mathrm{GHz},2\,\mathrm{GHz}\}$ only for a negligible part of the trajectory no distance measurements are associated.
However, at $BW=500\,\mathrm{MHz}$ roughly $10\%$ and $20\%$ of trajectory positions yield no associated or estimated components.

Figure~\ref{fig:LOS} shows the estimated amplitudes (compensated for free space path loss) of the components associated to the LOS (Fig.~\ref{fig:LOS:amp:4GHz}, \ref{fig:LOS:amp:2GHz} and \ref{fig:LOS:amp:500MHz}) and histograms of the delay estimates along the trajectory relative to the ground truth LOS delay $\tau_{\mathrm{gt},1}^{(j)}$ for varying bandwidth (Fig.~\ref{fig:LOS:error:4GHz}, \ref{fig:LOS:error:2GHz} and \ref{fig:LOS:error:500MHz}).

The amplitudes are plotted over the angle of arrival (AOA) $\varphi$ of the LOS, with $\varphi=0~\mathrm{rad}$ pointing in the look direction of the person.
The AOAs are computed using the ground truth orientation $\psi_{\mathrm{gt},n}$ from the OTS.
An $8$th-order spline fit $b(\varphi)$ is included (solid black) for better comparison. 
With the agent positioned at the center torso, the shape of the $b(\varphi)$ is symmetric w.r.t. $\varphi=0~\mathrm{rad}$.
Even though fewer measurements are associated when reducing the bandwidth, the amplitudes exhibit a similar shape.

The LOS histograms show that at large bandwidth additional distance measurements after the true LOS are obtained, supporting the scattering model proposed in \cite{WildingPIMRC2020}, and showing that fewer scattered paths are observed when reducing the bandwidth.

\subsection{Tracking Performance}\label{sec:results:tracking}



The performance of the EKF is compared for bandwidth values of $BW=\{4\,\mathrm{GHz},2\,\mathrm{GHz},500\,\mathrm{MHz}\}$, using a cut-off distance $d_c=0.3~\mathrm{m}$ and bandwidth reduction factor of $\rho=10$.  
Based on effects observable from the associated LOS amplitudes (c.f. Fig.~\ref{fig:LOS} and Section~\ref{sec:results:analysis}) the FOV was chosen as a sector of $270~\mathrm{deg}$ centered at $\varphi=0~\mathrm{deg}$ with VAs or PAs being visible when the AOA is within the FOV sector.
Note that is assumed known to evaluate the FOV at time step $n$ the orientation.

To investigate the effect of the multipath-based tracking and FOV three configuration of the EKF are used for all bandwidth values: (I) LOS-only with FOV, (II) multipath-based with FOV and (III) multipath-based without FOV.
Configuration (I) only allows associations to the PA, while (II) allows associations to PAs and VAs within the FOV. 
Configuration (III) assumes a uniform FOV and allows associations to all PAs and VAs geometrically visible at the current position, thus needing no knowledge of the orientation.
The position error $e_{\mathrm{pos},n} = \|\vm{p}_{\mathrm{gt},n}-\vm{p}_n\|$ for the different configurations is shown as a CDF in Fig.~\ref{fig:EKF:CDF} and along the trajectory in Fig.~\ref{fig:EKF:error}.

The CDF curves in Fig.~\ref{fig:EKF:CDF} show that choosing a larger bandwidth increases the accuracy when using multipath-based tracking and the FOV.
The LOS-only cases are strongly affected by shadowing introduced by the user as at least two PAs are necessary to continue following the track. 

The position error curves in Fig.~\ref{fig:EKF:error} show that selecting $BW=\{4\,\mathrm{GHz},2\,\mathrm{GHz}\}$ allows to navigate the full track with high accuracy when using multipath-based tracking and the FOV (II).
Using only multipath-based tracking (III) shows a similar performance until trajectory position $n=400$ (at $BW=4\,\mathrm{GHz}$) and $n=650$ (at $BW=4\,\mathrm{GHz}$) where the EKF diverges from the track.
While following the track the position error is below $0.2~\mathrm{m}$ for configurations (II) and (III) and for $BW=\{4\,\mathrm{GHz},2\,\mathrm{GHz}\}$.
For the low bandwidth of $BW=500\,\mathrm{MHz}$ the EKF quickly deviates at around $n=100$, due an insufficient number of associated measurements.

\section{Conclusion and Future Work}\label{sec:conclusion}

In this work we have shown that robust localization in realistic off-body channels is only possible when multipath components and the shadowing effects of the human body are considered. 
The proposed multipath-based localization and tracking algorithm extends \cite{MeissnerWCL2014} by a FOV considering the shadowing effects of the human body. 
With this model the visibilities of MPCs can be well predicted, leading to an improvement of the data association robustness and to a significant increase of position accuracy. 
However, the experimental results have shown that the number of detected MPCs in the radio signal decreases with decreased signal bandwidth making the tracking very unstable.

To overcome these issues, future work will focus on BP-based SLAM algorithms that incorporate probabilistic data association and adaptive detection probabilities by exploiting amplitude information \cite{LeitingerICC2019} or a discrete detection probability state \cite{LeitingerICC2017}. 
Such algorithms can inherently cope with shadowing and dispersive effects induced by the human body as well as geometrical visibilies of map features (VAs). 
Furthermore, these algorithms can detect and exploit features with very low component SNR \cite{BPChannelTracker_ArXiv2021}.

%

\section*{Acknowledgment}

This research was partly funded by the Austrian Research Promotion Agency (FFG) within the project UBSmart (project number: 859475) and by Graz University of Technology within the Lead Project "Dependable Things".
The financial support by the Christian Doppler Research Association, the Austrian Federal Ministry for Digital and Economic Affairs and the National Foundation for Research, Technology and Development is gratefully acknowledged.

\bibliographystyle{IEEEtran}
\bibliography{IEEEabrv,./EuCAP}

\begin{thebibliography}{10}
\providecommand{\url}[1]{#1}
\csname url@samestyle\endcsname
\providecommand{\newblock}{\relax}
\providecommand{\bibinfo}[2]{#2}
\providecommand{\BIBentrySTDinterwordspacing}{\spaceskip=0pt\relax}
\providecommand{\BIBentryALTinterwordstretchfactor}{4}
\providecommand{\BIBentryALTinterwordspacing}{\spaceskip=\fontdimen2\font plus
\BIBentryALTinterwordstretchfactor\fontdimen3\font minus
  \fontdimen4\font\relax}
\providecommand{\BIBforeignlanguage}[2]{{%
\expandafter\ifx\csname l@#1\endcsname\relax
\typeout{** WARNING: IEEEtran.bst: No hyphenation pattern has been}%
\typeout{** loaded for the language `#1'. Using the pattern for}%
\typeout{** the default language instead.}%
\else
\language=\csname l@#1\endcsname
\fi
#2}}
\providecommand{\BIBdecl}{\relax}
\BIBdecl

\bibitem{TarMupRauSloSveWym:SPM2014_LocationAware5G}
R.~Di~Taranto, S.~Muppirisetty, R.~Raulefs, D.~Slock, T.~Svensson, and
  H.~Wymeersch, ``Location-aware communications for 5{G} networks: {How}
  location information can improve scalability, latency, and robustness of
  5{G},'' \emph{IEEE Signal Process. Mag.}, vol.~31, no.~6, pp. 102--112, Nov.
  2014.

\bibitem{BressonTIV2017}
G.~Bresson, Z.~Alsayed, L.~Yu, and S.~Glaser, ``Simultaneous localization and
  mapping: {A} survey of current trends in autonomous driving,'' \emph{IEEE
  Trans. Intell. Veh.}, vol.~2, no.~3, pp. 194--220, Sept. 2017.

\bibitem{WinSPM2018}
M.~Z. Win, F.~Meyer, Z.~Liu, W.~Dai, S.~Bartoletti, and A.~Conti, ``Efficient
  multisensor localization for the internet of things: Exploring a new class of
  scalable localization algorithms,'' \emph{{IEEE} Signal Process. Mag.},
  vol.~35, no.~5, pp. 153--167, Sept. 2018.

\bibitem{MeissnerWCL2014}
P.~Meissner, E.~Leitinger, and K.~Witrisal, ``{UWB} for robust indoor tracking:
  Weighting of multipath components for efficient estimation,'' \emph{IEEE
  Wireless Comm. Lett.}, vol.~3, no.~5, pp. 501--504, Oct. 2014.

\bibitem{LeitingerGNSS2016}
E.~Leitinger, F.~Meyer, P.~Meissner, K.~Witrisal, and F.~Hlawatsch, ``Belief
  propagation based joint probabilistic data association for multipath-assisted
  indoor navigation and tracking,'' in \emph{Proc. ICL-GNSS-16}, Barcelona,
  Spain, June 2016, pp. 1--6.

\bibitem{WitrisalSPM2016}
K.~Witrisal, P.~Meissner, E.~Leitinger, Y.~Shen, C.~Gustafson, F.~Tufvesson,
  K.~Haneda, D.~Dardari, A.~F. Molisch, A.~Conti, and M.~Z. Win,
  ``High-accuracy localization for assisted living: 5g systems will turn
  multipath channels from foe to friend,'' \emph{{IEEE} Signal Process. Mag.},
  vol.~33, no.~2, pp. 59--70, 2016.

\bibitem{SchMarGenSanFie:AWPL2021_MultipathEnhancedLoc}
M.~Schmidhammer, C.~Gentner, S.~Sand, and U.-C. Fiebig, ``Multipath-enhanced
  device-free localization in wideband wireless networks,'' \emph{{IEEE}
  Antennas Wireless Propag. Lett.}, vol.~20, no.~4, pp. 453--457, Jan. 2021.

\bibitem{LiJSTSP2013}
L.~Li and J.~L. Krolik, ``Simultaneous target and multipath positioning,''
  \emph{{IEEE} J. Sel. Topics Signal Process.}, vol.~8, no.~1, pp. 153--165,
  2013.

\bibitem{GentnerTWC2016}
C.~Gentner, W.~Jost, T.and~Wang, S.~Zhang, A.~Dammann, and U.~C. Fiebig,
  ``Multipath assisted positioning with simultaneous localization and
  mapping,'' \emph{{IEEE} Trans. Wireless Commun.}, vol.~15, no.~9, pp.
  6104--6117, Sept. 2016.

\bibitem{LeitingerICC2017}
E.~Leitinger, F.~Meyer, F.~Tufvesson, and K.~Witrisal, ``Factor graph based
  simultaneous localization and mapping using multipath channel information,''
  in \emph{Proc. IEEE ICCW-17}, Paris, France, May 2017, pp. 652--658.

\bibitem{LeitingerTWC2019}
E.~{Leitinger}, F.~{Meyer}, F.~{Hlawatsch}, K.~{Witrisal}, F.~{Tufvesson}, and
  M.~Z. {Win}, ``A belief propagation algorithm for multipath-based {SLAM},''
  \emph{{IEEE} Trans. Wireless Commun.}, vol.~18, no.~12, pp. 5613--5629, Dec.
  2019.

\bibitem{LeitingerICC2019}
E.~{Leitinger}, S.~{Grebien}, and K.~{Witrisal}, ``Multipath-based {SLAM}
  exploiting {AoA} and amplitude information,'' in \emph{Proc. IEEE ICCW-19},
  Shanghai, China, May 2019, pp. 1--7.

\bibitem{MendrzikJSTSP2019}
R.~{Mendrzik}, F.~{Meyer}, G.~{Bauch}, and M.~Z. {Win}, ``Enabling situational
  awareness in millimeter wave massive {MIMO} systems,'' \emph{{IEEE} J. Sel.
  Topics Signal Process.}, vol.~13, no.~5, pp. 1196--1211, Sep. 2019.

\bibitem{LeitingerAsilomar2020_DataFusion}
E.~Leitinger and F.~Meyer, ``Data fusion for multipath-based {SLAM},'' in
  \emph{Proc. Asilomar-20}, Pacifc Grove, CA, USA, Oct. 2020.

\bibitem{LeitingerJSAC2015}
E.~Leitinger, P.~Meissner, C.~Rudisser, G.~Dumphart, and K.~Witrisal,
  ``Evaluation of position-related information in multipath components for
  indoor positioning,'' \emph{{IEEE} J. Sel. Areas Commun.}, vol.~33, no.~11,
  pp. 2313--2328, Nov. 2015.

\bibitem{WitrisalWCL2016}
K.~Witrisal, E.~Leitinger, S.~Hinteregger, and P.~Meissner, ``Bandwidth scaling
  and diversity gain for ranging and positioning in dense multipath channels,''
  \emph{IEEE Wireless Commun. Lett.}, vol.~5, no.~4, pp. 396--399, 2016.

\bibitem{hinteregger2016}
S.~Hinteregger, E.~Leitinger, P.~Meissner, J.~Kulmer, and K.~Witrisal,
  ``Bandwidth dependence of the ranging error variance in dense multipath,'' in
  \emph{Signal Processing Conference (EUSIPCO), 2016 24th European}.\hskip 1em
  plus 0.5em minus 0.4em\relax IEEE, 2016, pp. 733--737.

\bibitem{MendrzikTWC2019}
R.~Mendrzik, H.~Wymeersch, G.~Bauch, and Z.~Abu-Shaban, ``Harnessing{ NLOS}
  components for position and orientation estimation in {5G} millimeter wave
  {MIMO},'' \emph{{IEEE} Trans. Wireless Commun.}, vol.~18, no.~1, pp. 93--107,
  Jan. 2019.

\bibitem{WildingPIMRC2020}
T.~{Wilding}, E.~{Leitinger}, U.~{Muehlmann}, and K.~{Witrisal}, ``Modeling
  human body influence in {UWB} channels,'' in \emph{2020 IEEE 31st Annual
  International Symposium on Personal, Indoor and Mobile Radio Communications},
  2020, pp. 1--6.

\bibitem{WildingEuCAP2021}
T.~Wilding, E.~Leitinger, U.~Muehlmann, and K.~Witrisal, ``Statistical modeling
  of the human body as an extended antenna,'' in \emph{2021 15th European
  Conference on Antennas and Propagation (EuCAP) (EuCAP 2021)}, D{\"u}sseldorf,
  Germany, Mar. 2021.

\bibitem{SantosTWC2010a}
T.~Santos, J.~Karedal, P.~Almers, F.~Tufvesson, and A.~F. Molisch, ``Modeling
  the ultra-wideband outdoor channel: Measurements and parameter extraction
  method,'' \emph{IEEE Transactions on Wireless Communications}, vol.~9, no.~1,
  pp. 282--290, 2010.

\bibitem{SchubertTAP2013}
F.~M. Schubert, M.~L. Jakobsen, and B.~H. Fleury, ``Non-stationary propagation
  model for scattering volumes with an application to the rural {LMS}
  channel,'' \emph{{IEEE} Trans. Antennas Propag.}, vol.~61, no.~5, pp.
  2817--2828, 2013.

\bibitem{FlordelisTWC2020}
J.~Flordelis, X.~Li, O.~Edfors, and F.~Tufvesson, ``Massive {MIMO} extensions
  to the {COST} 2100 channel model: Modeling and validation,'' \emph{{IEEE}
  Trans. Wireless Commun.}, vol.~19, no.~1, pp. 380--394, 2020.

\bibitem{ErtelJSAC1999}
R.~B. Ertel and J.~H. Reed, ``Angle and time of arrival statistics for circular
  and elliptical scattering models,'' \emph{{IEEE} J. Sel. Topics Signal
  Process.}, vol.~17, no.~11, pp. 1829--1840, 1999.

\bibitem{JakobsenIZS2012}
M.~L. Jakobsen, T.~Pedersen, and B.~H. Fleury, ``Analysis of the stochastic
  channel model by {S}aleh \& {V}alenzuela via the theory of point processes,''
  in \emph{22th International Zurich Seminar on Communications (IZS)}.\hskip
  1em plus 0.5em minus 0.4em\relax Eidgen{\"o}ssische Technische Hochschule
  Z{\"u}rich, 2012.

\bibitem{Ottersten1993}
B.~Ottersten, M.~Viberg, P.~Stoica, and A.~Nehorai, ``Exact and large sample
  maximum likelihood techniques for parameter estimation and detection in array
  processing,'' in \emph{Radar Array Processing}.\hskip 1em plus 0.5em minus
  0.4em\relax Springer, 1993, pp. 99--151.

\bibitem{NadlerTSP2011}
B.~{Nadler} and A.~{Kontorovich}, ``Model selection for sinusoids in noise:
  {S}tatistical analysis and a new penalty term,'' \emph{{IEEE} Trans. Signal
  Process.}, vol.~59, no.~4, pp. 1333--1345, Apr. 2011.

\bibitem{LeitingerAsilomar2020}
E.~Leitinger, S.~Grebien, B.~H. Fleury, and K.~Witrisal, ``Detection and
  estimation of a spectral line in {MIMO} systems,'' in \emph{Proc.
  Asilomar-20}, Pacifc Grove, CA, USA, Oct. 2020, pp. 1090--1095.

\bibitem{SchuhmacherTSP2008}
D.~Schuhmacher, B.-T. Vo, and B.-N. Vo, ``A consistent metric for performance
  evaluation of multi-object filters,'' \emph{IEEE transactions on signal
  processing}, vol.~56, no.~8, pp. 3447--3457, 2008.

\bibitem{BPChannelTracker_ArXiv2021}
X.~{Li}, E.~{Leitinger}, A.~{Venus}, and F.~{Tufvesson}, ``Sequential detection
  and estimation of multipath channel parameters using belief propagation,''
  2021, arXiv:2109.05623.

\end{thebibliography}

\end{document}